\documentclass[letterpaper, 10 pt, conference]{ieeeconf}  

\IEEEoverridecommandlockouts                              

\title{\LARGE \bf
Optimal Utility Design with Arbitrary Information Networks
}

\author{Vartika Singh, Will Wesley, and Philip N. Brown
\thanks{*This material is based upon work supported by the Air Force Office of Scientific Research under award number FA9550-23-1-0171.}
\thanks{The authors are with the University of Colorado Colorado Springs, USA {\tt \{vsingh,wwesley,pbrown2\}@uccs.edu}}%
}

\usepackage{amsmath,amssymb}
\usepackage{enumerate}
\usepackage{xcolor}
\usepackage{mathtools}
\usepackage{wrapfig}
\newtheorem{theorem}{Theorem}
\newtheorem{remark}{Remark}

\newtheorem{lemma}{Lemma}

\newtheorem{definition}{Definition}
\newcommand{\A}{{\cal A}}
\newcommand{\f}{\Tilde{f}}
\newcommand{\hide}[1]{}

\renewcommand{\aa}{{\bf a}}

\newcommand{\I}{{\cal I}}
\newcommand{\R}{{\cal R}}
\newcommand{\C}{{\cal C}}
\renewcommand{\O}{{\cal O}}
\newcommand{\N}{{\cal N}}
\newcommand{\poa}{\mbox{PoA}(f,w,\C,\N)}
\newcommand{\poagen}{\mbox{PoA}(\{f_i\},w,\N)}
\newcommand{\GG}{{\cal G}_{f,w}^{(\C,\N)}}
\newcommand{\hGG}{\hat{\cal G}_{f,w}^{(\C,\N)}}
\newcommand{\hatGG}{\hGG}
\newcommand{\aopt}{a^{\mbox{opt}}}
\newcommand{\ane}{a^{\mbox{ne}}}

\newcommand{\gentuple}{(a_1,x_1,b_1),\dots,(a_k,x_k,b_k)}
\newcommand{\ttuple}{t_1\dots,t_k}
\newcommand{\tltuple}{\Tilde{t}_1\dots,\Tilde{t}_k}
\newcommand{\eop}{{\hfill $\blacksquare$}}
\newcommand{\n}{\Tilde{n}}

\newcommand{\fbl}{f_{bl}}
\newcommand{\fnbl}{f_{nbl}}
\newcommand{\TR}[2]{#2}                    
\begin{document}

\maketitle
\thispagestyle{empty}
\pagestyle{empty}

\begin{abstract}
We consider multi-agent systems with general information networks where an agent  may only  observe  a subset of other agents. A system designer assigns local utility functions to the agents guiding their actions towards  an outcome which determines the value of a given system objective. The aim is to design these local utility functions such that the Price of Anarchy (PoA), which equals the ratio of system objective  at worst possible outcome to that at the optimal, is maximized. Towards this, we first develop a linear program (LP) that characterizes the PoA for any utility design and any information network. This leads to another LP that optimizes the PoA and derives the optimal utility design. Our work substantially generalizes existing approaches to the utility design problem. We also numerically show the robustness of  proposed framework against unanticipated communication failures.
\end{abstract}

\section{Introduction}
Multi-agent systems have a  variety of  applications ranging from  air traffic control \cite{atc1}-\cite{atc2}, to the use of robotic networks in post disaster environments \cite{rn1}-\cite{rn3}, airport security \cite{airport}, transportation networks \cite{tn2} and  medical sciences \cite{ms1}-\cite{ms2} etc.  These systems require coordination among agents  to achieve a desired system objective. Centralized control may be infeasible or expensive due to demand for information, thus a game theoretic approach that allows for distributed control is a topic of interest among researchers \cite{Marden}-\cite{Ramaswamy}.

A system designer assigns local utility functions to agents  that guide their actions and determine the value of a given system objective. The utility functions are designed to depend  only 
on the local information available to the agents. The performance of such systems is measured using the well known Price of Anarchy metric \cite{PoA}, which compares the system objective at the worst Nash equilibrium of the game to the optimal system objective. 
A growing literature analyses the price of anarchy (PoA) for classes of games resulting from various utility functions; e.g., \cite{vetta}-\cite{Ramaswamy}. 

An important aspect of such problems is the \textit{utility design}, where the system designer aims to design local utility functions that improve the PoA of the resulting game. For example,  \cite{Gairing} analyses \textit{set covering games} and provides  local utility functions that result in a PoA of $1-1/e$ (where $e$ is Euler's constant). Further, \cite{Marden} provides a linear program (LP)  that derives the PoA-optimizing utility design for a wide class of resource allocation problems, whereas \cite{Marden2} specializes this result to classes of sub-modular, super-modular and set covering games. However,  \cite{Marden} and \cite{Marden2} both consider the case where agents can observe the actions of all agents while making a decision. 

Many times, the agents may not have access to the actions of others possibly due to a communication failure, jamming by an adversary, information sharing could be expensive, etc. Such problems have been considered in \cite{Grimsman2018}-\cite{Josh2023} for specific utility designs and system objectives. For example, \cite{Grimsman2018} considers sub-modular system objective and a greedy approach for agents and provides upper and lower bound on PoA; \cite{Grimsman2020} and \cite{Grimsman2022} consider  \textit{valid utility games} that result from sub-modular objective and specific utility design, and derive the corresponding PoA.

In this work, we consider a wide class of multi-agent problems and arbitrary information networks   where the agents may be blind, isolated or may have access only to the actions of a subset of other agents. Inspired by \cite{Marden}, we address the utility design problem using an LP approach. 
 The developed framework is applicable to a variety of problems, and has set covering games, valid utility games and the framework of \cite{Marden} as special cases.  Our main contributions are as follows: (i) we first develop an LP that characterizes the \textit{exact  PoA} for any fixed utility design, information network and system objective; and (ii) we then develop another LP for any information network and system objective \textit{that derives optimal utility design resulting in optimal PoA}. 

 
For set covering games and an information network with blind agents, \cite{Grimsman2020} provides the PoA for the well-known  marginal contribution utility design. We numerically solve our LP deriving  optimal PoA, and show that  the optimal PoA equals the PoA at marginal contribution. Thus marginal contribution utility design is optimal for a network with blind agents. We then consider sub-modular system objective and the case of unanticipated communication failures. The designer may not know the exact information network that will be realized when designing the utility functions for various agents, and hence can no longer optimize the utility design for specific information network. We numerically show that the optimal utility design for a network with full information (complete graph) is robust and results in near optimal PoA even when some or all agents become blind or isolated. This depicts the robustness of the proposed framework against unanticipated communication failures. 

\section{Problem Description}
Let $\R =\{r_1, \dots,r_m\}$ be a finite set of resources where every resource $r\in \R$ is associated with a value $v_r\ge 0$. Let $N = \{1,\dots,n\}$ be a finite set of agents. Any agent $i \in N$ has an action set $\A_i \subseteq 2^\R$ and can choose resources as described by $\A_i$. The welfare generated at a resource $r$ depends on $v_r$ and the number of agents selecting $r$. Let $\A:=\A_1\times\dots\times \A_n$ denote the set of action profiles of the agents.  For $\aa = (a_1,\dots,a_n) \in \A $, define $|\aa|_r$ to be the number of agents selecting resource $r$ in action profile $\aa$. Then the  total welfare generated under action profile $\aa$, $W:\A \to \mathbb{R}$ is given by,
\begin{eqnarray}\label{eqn_syst_obj}
    W(\aa) &=& \sum_{r \in \cup_i a_i} v_r w(|\aa|_r),
\end{eqnarray}where \emph{basis function} $w(|\aa|_r)>0$  scales the value of a resource $r$ depending on the number of agents selecting it.  The task of the system designer is to assign local utility functions to the agents that guide their actions towards an action profile that maximizes the  system objective \eqref{eqn_syst_obj}.

\subsection{Local Utility Functions and Resulting Class of Games}
This work considers general information networks, where the information available to the agents can be described using an information graph  $\N:=(\N_1,\dots,\N_n)$, where $\N_i$ is the set of agents whose actions agent $i$ can observe (naturally, $i\in \N_i$). \hide{Thus,  the local utility functions should only depend on the local information available to respective agents.}  Let $[p]$ represent the set $\{1,\dots,p\}$ for any positive integer $p$ and $\{0,[p]\}$ represent $\{0,1,\dots,p\}$.  Let $|\aa|^{\N_i}_r$ be the number of agents in  $\N_i$ that select resource $r$, and $f_i: [n] \to \mathbb{R}$ be a mechanism that scales the perceived value of a resource if it is chosen  by $|\aa|^{\N_i}_r$ number of agents. Then,  the  local utility functions can be defined as,
\begin{eqnarray}\label{eqn_util}
     U_i(\aa) &=& \sum_{r \in  a_i} v_r f_i(|\aa|^{\N_i}_r), \hspace{5mm} \mbox{ for } i \in N.
\end{eqnarray}This leads to a game $G = \langle \R,w, N, \{\A_i\}_{i \in N},\N, \{f_i\}_{i\in N} \rangle$. We focus on the solution concept of pure strategy Nash equilibrium (NE) throughout this paper. For any $\aa$, let $(\Tilde{a}_i,\aa_{-i})$ represent the action profile where agent $i$ unilaterally deviates to $\Tilde{a}_i$ from $a_i$. Then NE is defined as
\begin{definition}[Nash Equilibrium \cite{Nash}]
   For any $G$, an action profile $\aa$ is a pure strategy Nash equilibrium if $U_i(\aa) \ge U_i(\Tilde{a}_i,\aa_{-i})$ for all $\Tilde{a}_i \in \A_i$ and all  $i \in N$.
\end{definition}

Based on some learning rule  \cite{learning}, the agents arrive to an NE represented by $\ane$, and the system objective \eqref{eqn_syst_obj} equals $W(\ane)$. The aim of the system designer is to choose utility generating mechanism $\{f_i\}$ such that $W(\ane)$ is `close' to $W(\aopt)$ for all NE of the resulting game, where $\aopt$ is the optimizer of \eqref{eqn_syst_obj}.

In general, the exact resource set $\R$ and exact action sets of the agents $\{\A_i\}$ may not be known \cite{Marden}. The system designer may only know the number of agents $n$, basis function $w(\cdot)$ and information graph $\N$, and choose mechanism $f=\{f_i\}_{i\in N}$ only based on this information. Let ${\cal G}_{f,w}^\N$ be defined to be a class of resource allocation problems/games that have
\begin{enumerate}[\textbf{G}.1]
    \item $n$ agents,  any possible $\R$, any possible $\{\A_i\}$,
    \item system objective defined as \eqref{eqn_syst_obj} with $w(\cdot)$ as the basis function, and  $W(\aa)>0$ for some $\aa \in \A$,
    \item local utilities as in \eqref{eqn_util} with $\{f_i\}_{i\in N}$ as  utility generating mechanism and  $\N$ as the information graph. 
\end{enumerate}Then, the performance of $f$ can be measured using Price of Anarchy (PoA) metric (see \cite{PoA}), defined as,
\begin{eqnarray}\label{eqn_def_poa}
  \poagen = \inf_{G \in {\cal G}_{f,w}^\N}\left(\frac{\min_{\aa \in NE(G)} W(\aa)}{\max_{\aa \in \A}W(\aa)}\right),
\end{eqnarray}where $NE(G)$ is the set of all NE of $G\in {\cal G}_{f,w}^\N$. To optimize PoA, one first needs to characterize \eqref{eqn_def_poa} for any $\{f_i\}$. In the next section, we  present required structure which will facilitate the same.

\subsection{Information Graph and Classes}
\begin{figure}
    \centering  
    \includegraphics[trim ={8cm 9.1cm 9cm 8cm},clip,scale=0.45]{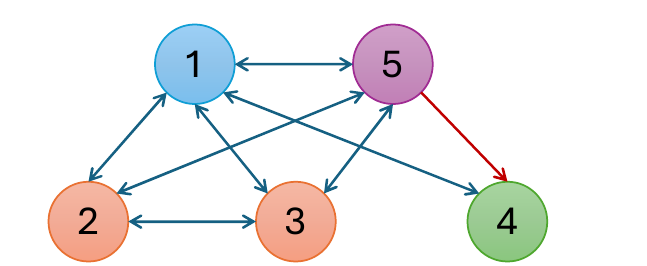}
    \caption{A general network with information available to various agents given by $\N_1 = \N_5 = \{1,2,3,4,5\}$, $\N_2 =\N_3= \{1,2,3,5\}$ and $ \N_4 =\{1,4\}$. Agent 1 and 5 observe the same set of agents, but 1 can be observed by 4 whereas 5 cannot be, thus actions of 1 affect 4 differently than actions of 5, hence, 1 and 5 are not similar. On the other hand  2 and 3 are similar. The agents can be partitioned as $\C_1 =\{1\}$, $\C_2 = \{2,3\}$, $\C_3 = \{4\}$ and $\C_4 =\{5\}$ based on similarity.  Then $\N_{\C_1}= \N_{\C_4}=\C_1\cup\C_2\cup \C_3\cup \C_4$, $\N_{\C_2} = \C_1 \cup \C_2\cup \C_4$ and $\N_{\C_3} = \C_1\cup \C_3$, and \textbf{C}.1-\textbf{C}.3 are satisfied. 
}
    \label{fig_networks}
\end{figure}
 One can utilize the given information graph $\N$ to  partition the set of agents to  classes of agents with `similar attributes'. We say two agents $i$ and $j$ are similar if (i) they can observe the same set of agents, that is, $\N_i=\N_j$, and (ii) if they can be observed by the same set of agents, that is,  $\{\N_l: i \in \N_l,\ l\in N\}=\{\N_l: j \in \N_l,\ l\in N\}$. The motivation behind such a partition stems from the fact that  similar agents affect the game in a similar manner, thus the system designer can assign them the same local utility functions.  Let $\C_j$ represent a class of agents and $\N_{\C_j}$ be the set of agents whose actions agents in class $\C_j$ can observe (see Figure \ref{fig_networks}).  
%
%
A partition of $N$ such that similar agents are assigned to the same class is accomplished by a class division where,
\begin{enumerate}[\textbf{C}.1]
    \item every agent is assigned to exactly one class,
    \item all the agents in a class $\C_j$ observe everyone in $\C_j$,
    \item the agents in  $\C_j$ observe everyone in  classes $m_1,\dots,m_{j}$ and not partial classes, that is, $\N_{\C_j} = \C_{m_1}\cup\dots\cup\C_{m_j}$.
\end{enumerate}Observe that \textbf{C}.2 along with \textbf{C}.3 implies that $\N_i =\N_{i'}$ for $i,i'\in \C_j$. Further, \textbf{C}.3 also implies that all the agents in a class are observed by the same set of agents\footnote{It is important to note that the partitioning of any network according to \textbf{C}.1-\textbf{C}.3 is not restrictive. Indeed,  any network can be partitioned by  making singleton classes $\C_i=\{i\}$ for $i\in N$.}.  \textit{With slight abuse of notation, from this point onward, we represent $\N_{\C_j}$, the information available to class $\C_j$,  as $\N_j$.} Then, any network can be completely specified using $(\C,\N)$ where $\C=\{\C_1,\dots \C_k\}$ is  the set of classes  of agents for some $k \le n$, and $\N=\{\N_1,\dots,\N_k\}$ represents the information available to various classes.
 
\subsection{Price of Anarchy for Network $(\C,\N)$} 
Let $f_j:[|\N_{j}|]\to \mathbb{R}$ be the utility generating mechanism for class $\C_j$.
 Let $|\aa|^{\N_j}_r$ be the number of agents in $\N_j$ selecting resource $r$ in action profile $\aa$. Then,
the utility of an agent $i \in \C_j$ equals
\begin{eqnarray}\label{eqn_util_gen}
     U_i(a_i,a_{-i})&=& \sum_{r \in  a_i} v_r f_j(|\aa|^{\N_j}_r). 
\end{eqnarray} Let ${\cal G}_{f,w}^{(\C,\N)}$ be defined to be a class of problems/games that satisfy $\textbf{G.1}-\textbf{G}.2$, and  have
\begin{enumerate}[\textbf{G}.1]
\setcounter{enumi}{3}
   \item local utilities given by \eqref{eqn_util_gen} with $\{f_j\}_{j\in[k]}$ as  utility generating mechanism and  $(\C,\N)$ as the network, where $\C =\{\C_1,\dots,\C_k\}$ and $\N=\{\N_1,\dots,\N_k\}$ for $k \le n$. 
\end{enumerate}Let $f$  represent $(f_1,\dots,f_k)$.  Then, the PoA for a communication denied network $(\C,\N)$ with basis function $w$ and  utility generating mechanism $f$  is given by 
\begin{eqnarray}\label{eqn_poa_gen}
    \poa= \inf_{G \in \GG}\left(\frac{\min_{\aa \in NE(G)} W(\aa)}{\max_{\aa \in \A}W(\aa)}\right).
\end{eqnarray}
The authors in \cite{Marden} consider the full information case where all the agents observe actions of all the other agents and provide a linear program that derives the optimal utility generating mechanism and optimal PoA (that is, the special case of our problem with $\C=\{\C_1\}$ and ${\cal N}_1=N$). In this work, we aim to derive an optimal utility generating mechanism $f^{opt}=\{f^{opt}_j\}_{j\in[k]}$ that optimizes the PoA defined in \eqref{eqn_poa_gen} for any general network $(\C,\N)$ and basis function $w$, 
\begin{eqnarray}\label{eqn_poa_opt}
   f^{opt}&=& \arg\max_f \ \poa.
\end{eqnarray}

\section{Linear Program for Price of Anarchy}In this section, we develop a linear program (LP) that computes the PoA of \eqref{eqn_poa_gen} for any given network $(\C,\N)$, basis function $w$ and fixed utility generating mechanism $f$.

Let  $\kappa_j$ be the number of agents in class $\C_j$, i.e., $\kappa_j = |C_j|\ge 1$ for $j \in [k]$. Let   $\O_j:=\{m_1,\dots,m_j\}$ represent the index of classes whose agents can be observed by agents of $\C_j$, that is,  $\N_j=\cup_{l \in \O_j} \C_l$ for $j \in [k]$. Define
%
\begin{eqnarray}\label{eqn_gen_I}
    \I &:= &\Big\{(\gentuple) : (a_j,x_j,b_j) \in \mathbb{N}^3_{\ge 0},\nonumber\\
    &&\hspace{1.8cm} 0 \ \le\  a_j+x_j+b_j \ \le\ \kappa_j \mbox{ for } j\in[k]\nonumber \\
    &&\hspace{1cm} \mbox{ and }\ 1 \ \le\  \sum_{j\in[k]} (a_j+x_j+b_j)  \le \ n\Big\}.
\end{eqnarray}For ease of notation, let $t$ be defined to be $(\ttuple)$ where $t_j=(a_j,x_j,b_j)$. Then a typical element of $\I$ is $t \in \I$.  Further, for any $t$ let
\begin{eqnarray}\label{eqn_short}
    A_t &:=& \sum_{j \in [k]} a_j + x_j, \hspace{2mm} B_t\ :=\ \sum_{j \in [k]} b_j + x_j \mbox{ and }\nonumber\\
     A_{t,j} &:=& \sum_{l \in \O_j} a_l + x_l \mbox{ for all } j \in [k].
\end{eqnarray} We now present our first main result (proof in Appendix).
\begin{theorem}[Primal LP to characterize PoA]\label{thm_primal_lp_gen}
For a given network $(\C,\N)$, basis function $w$ and utility generating mechanism $f=\{f_j\}_{j\in[k]}$,
\begin{enumerate}[(i)]
\item If $f_j(1)\le 0$ for any $j \in [k]$, then  $\poa=0$.
\item If $f_j(1)>0$ for all $j\in[k]$, then 
\begin{eqnarray}
    \poa &=& \frac{1}{W^*} \label{eq:poa_thm1}
\end{eqnarray}where $W^*$ is the finite solution of the following LP in the unknowns $\{\theta(t)\}$:
\end{enumerate}
\begin{eqnarray}\label{eqn_poa_primal_lp_gen}
   W^* &=& \max_{\theta(t)}\sum_{t\in \I} w(B_t) \theta(t) \hspace{5mm} \mbox{subject to} \nonumber\\
&& \hspace{-10mm} \sum_{t\in \I} [ a_j f_j(A_{t,j})-  b_j f_j(A_{t,j}+1)] \theta(t) \ge 0,\hspace{2mm} j \in [k]\nonumber\\
    && \hspace{-10mm}\sum_{t\in\I} w(A_t)\theta(t)=1,\nonumber\\
    && \hspace{-10mm} \theta(t) \ge 0 \ \hspace{2mm} \mbox{for all}\ \ t\in \I, \\
  \mbox{and} && \hspace{-10mm}\ f_j(0)= f_j(|\N_j|+1) = w(0) =0, \ \forall \ j\in[k].\nonumber
\end{eqnarray}

\end{theorem}
\vspace{2mm}

To the best of our knowledge, this is the first result that provides exact characterization of the PoA for any general
network $(\C,\N)$ and any utility generating mechanism $f$. Theorem \ref{thm_primal_lp_gen} is a non-trivial extension of  \cite[Theorem 2]{Marden}, which is a special case with only one class $\C_1$ and $\N_1=N$.
\begin{remark}
    The tightness of the bound~\eqref{eq:poa_thm1} in Theorem~\ref{thm_primal_lp_gen} guarantees that for every network, game instances exist which possess pure equilibria which respect the PoA bound.
    However, for some networks, it is possible to construct a game instance which has no pure NE. On such a game instance, Theorem~\ref{thm_primal_lp_gen} is uninformative as it only applies to pure equilibria. 
    Our ongoing work aims to address this gap.
\end{remark}

\subsection{Dual Linear Program to characterize PoA} This section presents the dual linear program for the LP in \eqref{eqn_poa_primal_lp_gen} and restricts the number of constraints to a smaller space $\I_\R$ in place of $\I$ defined as,
\begin{eqnarray}\label{eqn_gen_IR}
    \I_\R  \hspace{-2mm}&:= & \hspace{-2mm}\Big\{(\gentuple) : (a_j,x_j,b_j) \in \mathbb{N}^3_{\ge 0},\nonumber\\
    &&\ \ 0 \ \le\  a_j+x_j+b_j \ \le\ \kappa_j, \nonumber\\
    && a_j.x_j.b_j = 0\ \mbox{ or }\ a_j+x_j+b_j = \kappa_j\mbox{ for } j \in [k]\nonumber\\
    && \mbox{ and }\ 1 \ \le\  \sum_{j\in[k]} (a_j+x_j+b_j)  \le \ n\Big\}.
\end{eqnarray}Again, a typical element of $\I_\R$ is represented by $t\in\I_\R$. 
\begin{theorem}[Dual LP to characterize PoA]\label{thm_dual_lp_gen}
For a given network $(\C,\N)$, basis function $w$ and utility generating mechanism $f=\{f_j\}_{j\in[k]}$,
\begin{enumerate}[(i)]
 \item If $f_j(1)\le 0$ for any $j \in [k]$, then  $\poa=0$.
\item If $f_j(1)>0$ for all $j \in [k]$, then,
\begin{eqnarray*}
    \poa &=& \frac{1}{V^*}
\end{eqnarray*}where $V^*$ is the finite solution of the following (dual) LP in the unknowns $\mu,\lambda_1,\dots,\lambda_k$:
\end{enumerate}
\begin{eqnarray}\label{eqn_poa_dual_lp_gen}
V^* & =&  \min_{\lambda_1,\dots,\lambda_k \ge 0,\mu \in \mathbb{R}} \ \   \mu \hspace{5mm} \mbox{ subject to }\\
    && \hspace{-15mm} w(B_t) + \sum_{j \in [k]} \lambda_j  [ a_j f_j(A_{t,j})-  b_j f_j(A_{t,j}+1)]  \le  \mu  w(A_t)\nonumber\\
    && \hspace{4cm} \mbox{for all }t  \in \I_\R,\nonumber \\
    &&\hspace{-1.5cm} \mbox{and } f_j(0)= f_j(|\N_j|+1) = w(0) =0,\ \  \forall \ j\in[k]. \nonumber
\end{eqnarray}  

\end{theorem}
\vspace{2mm}


The characterization of $\poa$ using Theorem \ref{thm_dual_lp_gen} (proof in Appendix) paves the way towards finding an optimal mechanism $f^{opt}$ solving \eqref{eqn_poa_opt}. 
\section{Optimal Price of Anarchy} From Theorem \ref{thm_dual_lp_gen}, in order to find an optimal utility generating mechanism,  one needs to minimize $V^*$ of \eqref{eqn_poa_dual_lp_gen} which depends upon $f$. This appears to be a non-linear problem but since $\lambda_j$ and $f_j(\cdot)$ are always multiplied together, we have our next result.
\begin{theorem}[Optimal PoA for general network]\label{thm_opt_lp_gen}
For any basis function $w$ and  network $(\C,\N)$, the optimization problem \eqref{eqn_poa_opt} can be solved using the LP
\begin{eqnarray}\label{eqn_poa_opt_lp_gen}
    (f^{opt},\mu^{opt}) & \in &\arg  \min_{f,\mu} \   \mu\hspace{4mm} \mbox{ subject to}\nonumber\\
    && \hspace{-23mm} w(B_t) + \sum_{j \in [k]}  [ a_j f_j(A_{t,j})-  b_j f_j(A_{t,j}+1)]  \le  \mu  w(A_t) ,\nonumber\\
    && \hspace{3cm} \mbox{for all }t  \in \I_\R,\\
                  &&\hspace{-2.3cm}\mu\in \mathbb{R}, \ f_j(0)= f_j(|\N_j|+1) = w(0) =0,\ \forall \ j\in[k]. \nonumber
\end{eqnarray}The resulting optimal price of anarchy equals $1/\mu^{opt}$.
\end{theorem}
\vspace{2mm}
 \TR{}{The proof of Theorem \ref{thm_opt_lp_gen} follows in the exact same manner as the proof of \cite[Theorem 4]{Marden} by repeating the proof steps for all classes $\C_j$, $j\in[k]$. The proof is omitted in the view of space constraints.}
 
From Theorem \ref{thm_opt_lp_gen}, the system designer can design optimal local utility functions \eqref{eqn_util_gen} for any network $(\C,\N)$ and basis function $w$ using $f^{opt}$. This utility design guarantees a system objective always within a $1/\mu^{opt}$ factor of the optimal irrespective of the realised game $G$ or reached NE. 

\section{A Network with Blind Agents}  Consider a network where $\kappa$ agents for $\kappa \le n$ can only observe themselves, and thus are blind. The remaining $n-\kappa$ agents can observe the actions of all agents including the blind agents. \cite{Grimsman2020} provides the PoA for such networks for a  specific utility mechanism, marginal contribution. One can now use Theorem \ref{thm_opt_lp_gen} to derive an optimal mechanism.

Without loss of generality, say $i\in [\kappa]$ are blind agents, and $i=\{n-\kappa+1,\dots,n\}$ are non-blind agents.  One can partition the network using \textbf{C}.1-\textbf{C}.3 which leads to $(\kappa+1)$ classes. However, for this special network, we provide a refinement of Theorem \ref{thm_dual_lp_gen} and Theorem \ref{thm_opt_lp_gen} that has only two classes -- $\C_1$ for blind agents, and $\C_2$ for non-blind agents\footnote{Such a partition clearly violates $\textbf{C}.2$, and it becomes clear that  \textbf{C}.1-\textbf{C}.3 are not necessary assumptions for the framework in this work to be applicable. They only facilitate the assignment of utility functions to various classes.}. Since the blind agents can only observe themselves (thus only one agent), let $\fbl(1)$ be the utility generating mechanism assigned to all the blind agents; these agents do not need $\fbl(j)$ for $j\ge 2$. Let $\fnbl:[n]\to \mathbb{R}$ be the utility generating mechanism assigned to all the non-blind agents. Define $\I_\R$ as in \eqref{eqn_gen_IR} and
for each $t\in \I_\R$ define $A_t$, $B_t$ and $A_{t,2}$ as in \eqref{eqn_short}; note that one requires $\O_2 :=\{1,2\}$ to define $A_{t,2}$. Upon partitioning the set of agents in two classes, it is no longer possible to define $\O_1$ and thus $A_{t,1}$. However, our next theorem (proof in Appendix) shows that $A_{t,1}$ is not required to derive the PoA for $f=(\fbl,\fnbl)$ or  the optimal mechanism.

\begin{theorem}\label{thm_opt_lp_blind}
Consider a network with $\kappa$ blind agents and $n-\kappa$ non-blind agents for $\kappa \le n$, and let  $w \in \mathbb{R}^n_{>0}$ be a basis function,  then
\begin{enumerate}[(i)]
    \item If $\fbl(1) \le 0$ or $\fnbl(1) \le 0$, then PoA equals 0.
    \item If $\fbl(1)$ and $\fnbl(1) > 0$, then PoA equals $1/V^*$, where $V^*$ is the solution of following LP:
\begin{eqnarray}\label{eqn_poa_dual_lp_blind}
 && \hspace{-1cm} V^* \ =\  \min_{\lambda_1,\lambda_2,\mu} \   \mu\hspace{4mm} \mbox{ subject to}\nonumber\\
    && \hspace{-1cm} w(B_t)  + \lambda_1 (a_1-b_1) +\lambda_2 [ a_2 \fnbl(A_{t,2})\\
    && \hspace{10mm}-  b_2 \fnbl(A_{t,2}+1)] \ \le\  \mu  w(A_t) \hspace{2mm}\forall \ t  \in \I_\R,\nonumber \\
  &&\hspace{-1cm}\lambda_1,\lambda_2 \ge 0, \ \mu \in \mathbb{R},\hspace{2mm }\fnbl(0)= \fnbl(n+1) = w(0) =0. \nonumber
\end{eqnarray}
\item The optimal PoA is achieved by utility generating mechanism $(\fbl(1),\fnbl^{opt})$ where $\fbl(1)\in \mathbb{R}_{>0}$ and $\fnbl^{opt}$ is given by the solution of following LP
\begin{eqnarray}\label{eqn_poa_opt_lp_blind}
 && \hspace{-1cm} (\fnbl^{opt},\mu^{opt}) \ \in \  \arg\min_{\lambda,\fnbl,\mu} \   \mu\hspace{4mm} \mbox{ subject to}\nonumber\\
    && \hspace{-1cm} w(B_t)  + \lambda (a_1-b_1) + a_2 \fnbl(A_{t,2})\\
    && \hspace{10mm}-  b_2 \fnbl(A_{t,2}+1) \ \le\  \mu  w(A_t) \hspace{2mm}\forall \ t  \in \I_\R,\nonumber \\
  &&\hspace{-1cm}\lambda \ge 0, \ \mu \in \mathbb{R},\hspace{2mm }\fnbl(0)= \fnbl(n+1) = w(0) =0. \nonumber
\end{eqnarray}The resulting optimal price of anarchy equals $1/\mu^{opt}$.
\end{enumerate}
\end{theorem}
\vspace{2mm}


\subsection*{A network with isolated agents} Consider a network where $\kappa$ agents for $\kappa \le n$ can only observe themselves and cannot be observed by any other agent, thus are isolated. The remaining $n-\kappa$ agents can observe the actions of all agents except the isolated agents.  Then an exact similar result as Theorem \ref{thm_opt_lp_blind} holds with the only difference that utility generating mechanism for non-isolated agents is defined over set $[n-\kappa]$ in place of $[n]$. This result is omitted due to space constraints.  

\section{Numerical Results}
We present a few interesting examples related to networks with blind agents and with isolated agents. 

\subsection{Set Covering Games with Blind Agents}\label{sec_set_cover}
Set covering games are well studied in the literature for the full information case  \cite{Gairing}. Here, the system objective equals the sum of the values of the  resources selected by the agents. That is, $w(j)=1$ for all $j>1$ and $w(0)=0$ in \eqref{eqn_syst_obj}.  For a network with blind agents, \cite{Grimsman2020} provides the PoA for marginal contribution utilities which corresponds to utility generating mechanism $f^{mc}(j) = w(j)-w(j-1)$ with $w(0)=0$. Indeed, in this mechanism, the utility of any player upon selecting a resource equals the marginal gain in the system objective. Note that $f^{mc}$ for blind agents simply equals $w(1)$. The PoA for marginal contribution equals $\max(\frac{1}{1+\kappa},\frac{1}{n})$ when there are $\kappa \ge 1$ blind agents. Now we derive the optimal PoA using Theorem \ref{thm_opt_lp_blind}.

\begin{figure}
    \centering
    \includegraphics[trim={3cm 8.4cm 3cm 8cm},clip,scale = 0.5]{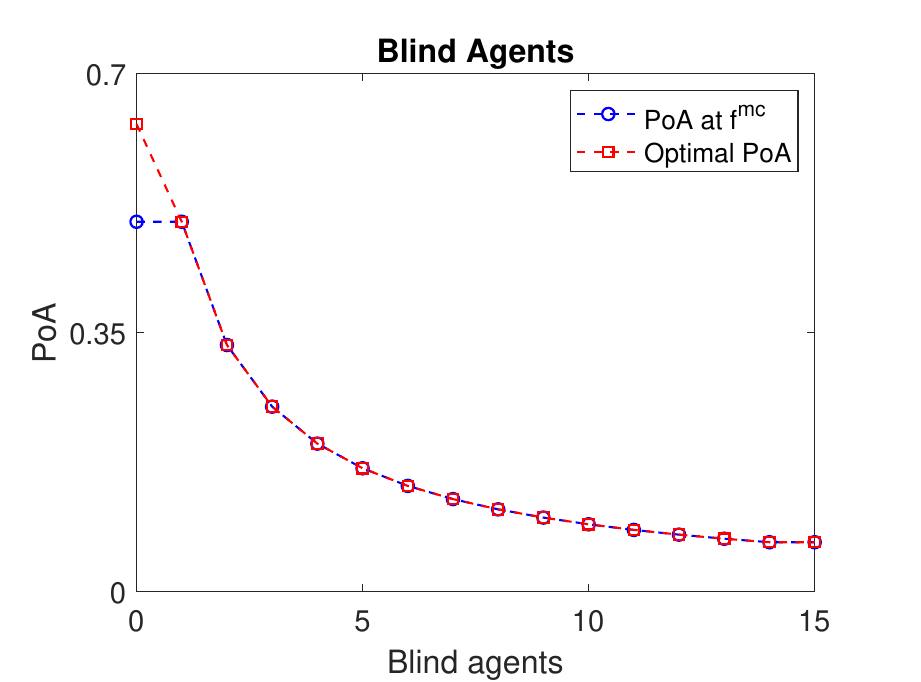}
    \caption{PoA at marginal contribution utility $f^{mc}$ and optimal PoA for set covering games of section \ref{sec_set_cover} as a function of number of blind agents in the network. The PoA at $f^{mc}$ matches with the optimal PoA if there is even one blind agent in the network, implying optimality of marginal contribution}
    \label{fig_Blind_set_cover_MC_PoA_vs_OptimalPoA}
\end{figure}

Consider a network of $n=15$ agents with $\kappa \in \{0,[n]\}$ blind agents and $n-\kappa$ non-blind agents. In Figure \ref{fig_Blind_set_cover_MC_PoA_vs_OptimalPoA}, we plot the optimal PoA obtained from Theorem \ref{thm_opt_lp_blind}  and PoA for marginal contribution $f^{mc}$, for various values of $\kappa$. Interestingly, the optimal PoA obtained from optimal mechanism matches exactly with that of marginal contribution mechanism for all $\kappa \ge 1$. 
Thus, the marginal-contribution utility mechanism is optimal for set covering games containing blind agents.

\begin{figure}
    \centering
    \begin{minipage}{0.45 \linewidth}
        \includegraphics[trim={3.5cm 8cm 4.2cm 8cm},clip,scale = 0.3]{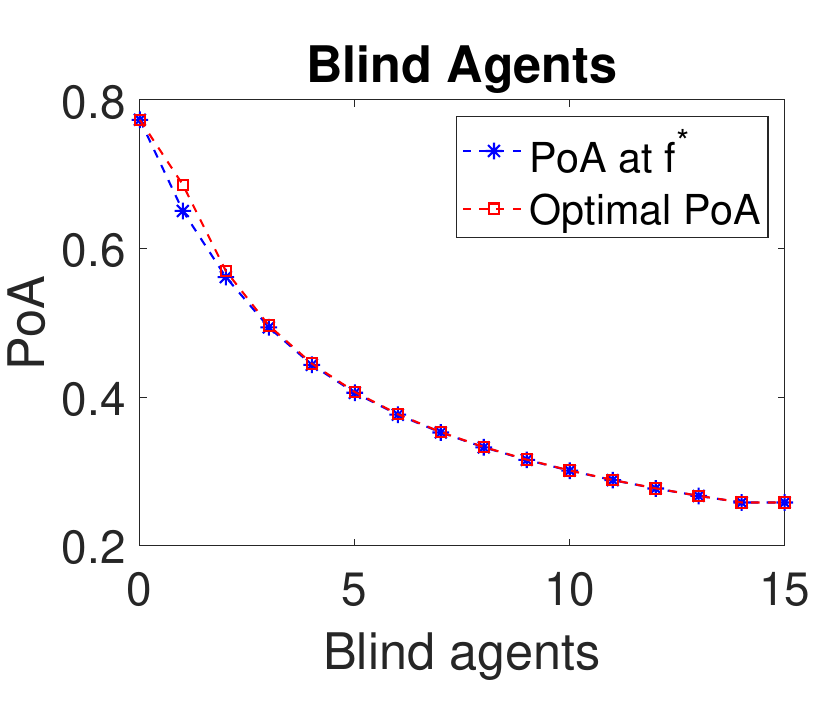}
    \end{minipage}\hspace{5mm}
    \begin{minipage}{0.45 \linewidth}
         \includegraphics[trim={3.5cm 8cm 4.2cm 8cm},clip,scale = 0.3]{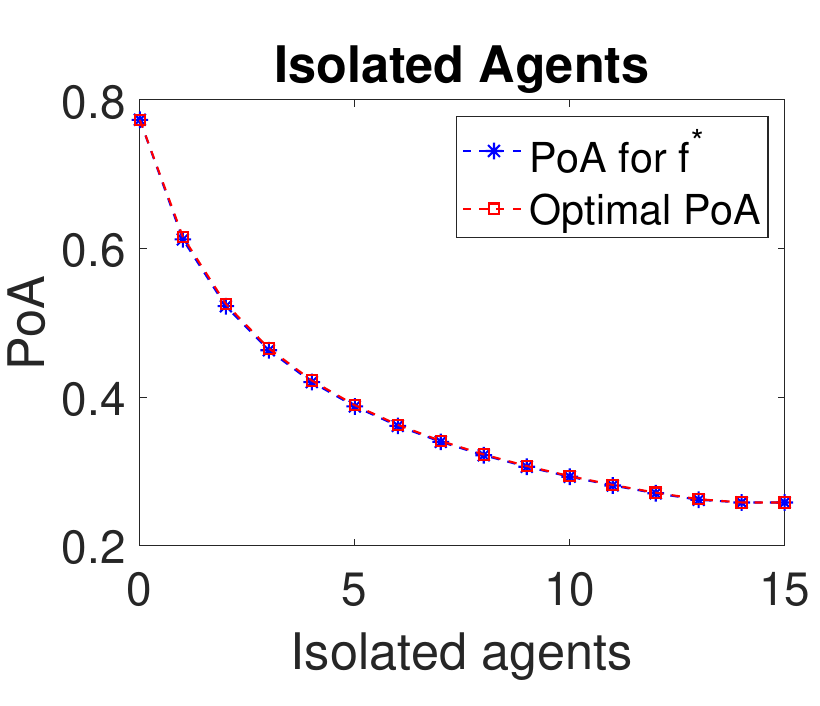}
    \end{minipage}
    \caption{PoA when agents keep using optimal utility mechanism corresponding to full information case $f^*$ even when there is a communication failure, compared to the optimal PoA  for sub-modular system objective. The mechanism $f^*$ is near optimal, thus robust against  communication failures.}
    \label{fig_blind_isolated_robustness}
\end{figure}

\subsection{Robustness of Optimal Mechanism} Consider a basis function defined as $w(j) = j^d$ with $d=0.5$, such a basis function leads to sub-modular system objective $W(\cdot)$ \cite{vetta}. Let the number of agents in the system be $n=15$, number of blind or isolated agents be $\kappa\in\{0,[n]\}$, and let $n-\kappa$ be the number of normal agents. Let $f^*$ be the optimal mechanism for the full information case (when $\kappa=0$) obtained from Theorem \ref{thm_opt_lp_gen} (or equivalently from \cite[Theorem 4]{Marden}). Say the system designer assigns $f^*$ to all the agents even if there is a communication failure leading to blind or isolated agents. In Figure \ref{fig_blind_isolated_robustness} we plot the PoA of utility generating mechanism $f^*$ and compare it with optimal PoA for (i) a network with blind agents in the left sub-figure, and (ii) for a network with isolated agents in the right sub-figure, for various values of $\kappa$. Interestingly, the mechanism $f^*$ is robust to communication failures. Indeed, for any network with $\kappa$ blind or isolated agents and $n-\kappa$ normal agents, the optimal PoA obtained from Theorem \ref{thm_opt_lp_blind} or Theorem \ref{thm_opt_lp_gen} is very close to that when agents keep using $f^*$. This observation suggests that in the event of uncertainty related to possible communication among the agents, it might be useful to assign the agents the (optimistic) utility mechanism $f^*$ optimal for full information case.

\hide{\begin{figure}
    \centering
    \includegraphics[trim={3cm 9.1cm 3cm 8cm},clip,scale = 0.5]{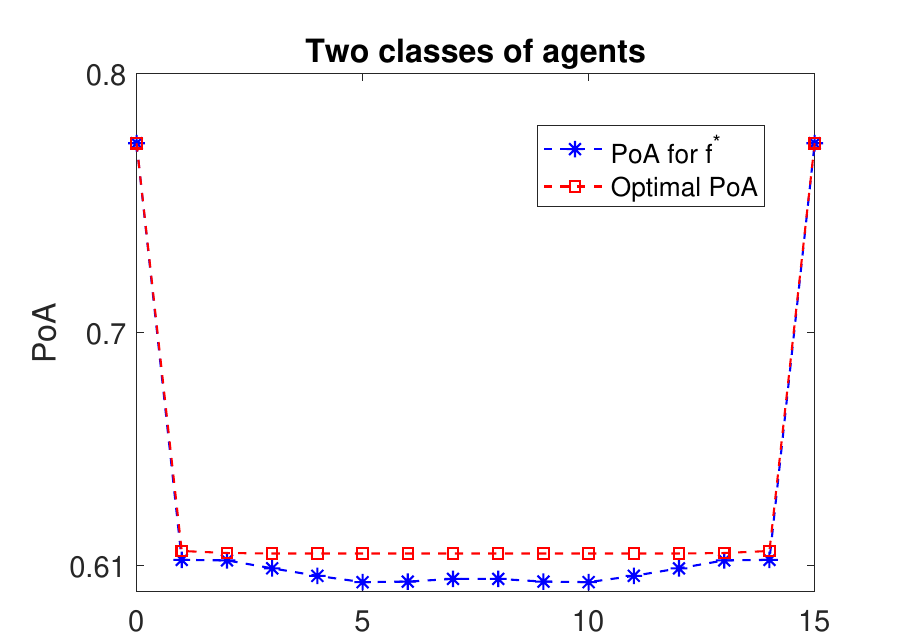}
    \caption{sub-modular -- repeating the optimal mechanism for full information case and plotting PoA of that vs respective optimal of isolated and blind}
    \label{fig:enter-label}
\end{figure}}


\section{Conclusions}
We address the utility design aspect in multi-agent coordination using game theory for general information networks. This paper presents two major results regarding the Price of Anarchy (PoA), which is the ratio of system objective at the worst case Nash equilibrium of the game resulting from a specific utility design, to the optimal system objective. For any information network among the agents, the first result provides a linear program (LP) that derives exact PoA for any choice of utility design. The second result  provides another LP that optimizes the PoA and derives the optimal utility design. Further, the numerical simulations suggest that the optimal utility design obtained for a network with complete graph is robust against communication failures. 

In future, we aim to perform a qualitative analysis of  the optimal PoA with respect to the number of edges in the information network. It would also be interesting to derive theoretical guarantees regarding the robustness of optimal utility design against unanticipated communication failures.  




\section*{APPENDIX}

\section*{Proof of Theorem \ref{thm_primal_lp_gen}}
\TR{The first part follows in the exact same manner as in \cite[Lemma 1]{Marden}. We prove part (ii) in four steps.  Step 1 and 2 follow in the exact same manner as in the proof of \cite[Theorem 2]{Marden}, which is reproduced for the sake of completeness.

\noindent\textbf{Step 1: }We prove that the $\poa$ in \eqref{eqn_poa_gen} computed over class $\GG$  is same as the price of anarchy computed over a reduced class of games $\hatGG \subset \GG$. Towards this, for every $G \in \GG$, consider a game $\hat{G}$ which is identical to $G$ in all the aspects except for the action sets of the agents. The action set of any agent $i$ in $\hat{G}$ only contains the action at worst NE of $G$ and that at optimal action profile, that is, $\A_i = \{\ane_i,\aopt_i\}$, and note that  the price of anarchy for $G$ and $\hat{G}$ are the same, 
\begin{eqnarray*}
  \frac{\min_{\aa \in NE(G)} W(\aa)}{\max_{\aa \in \A}W(\aa)} &=& \frac{ W(\ane)}{W(\aopt)}.
\end{eqnarray*}Thus, if $\hatGG$ is the set of games that contains one such $\hat{G}$ for all $G \in \GG$, then the PoA over $\hatGG$ is same as PoA over $\GG$.

\noindent\textbf{Step 2: }By \cite[Lemma 2]{Marden}, all the NE for $\hatGG$ games have strictly positive system objective at NE, $W(\ane)>0$. Thus each $\hat{G} \in \hatGG$, one can construct a game $\Tilde{G}$ which is same as $\hat{G}$ in all the aspects except for the value of resources. In $\Tilde{G}$, scale the value of any resource $r$ to $v_r/W(\ane)$ which in the original game $\hat{G}$ had a value of $v_r$. Thus the system objective at all allocations gets scaled by $W(\ane)$ and optimal and equilibrium action profiles still remain the same. Hence $\Tilde{G}$ has the same price of anarchy as $\hat{G}$. Thus the set of games $\{\Tilde{G} \in \hatGG: W(\ane)=1\}\subset \hatGG$ has the same price of anarchy as $\GG$ using the same reasoning as in step 1. Therefore, the price of anarchy can be computed as
\begin{eqnarray*}
    PoA(f,w,n) &=& \inf_{\hat{G} \in \hatGG} \hspace{2mm} \frac{1}{W(\aopt)}\\
    && \mbox{s.t } \hspace{5mm} U_i(\ane) \ \ge \ U_i(\aopt_i,\ane_{-i}) \hspace{5mm} \forall \ i\\
    && \hspace{1.1cm} W(\ane)\ = \ 1.
\end{eqnarray*}

\noindent\textbf{Step 3:}  From the last equation in step 2, observe that $PoA(f,w,n)= \frac{1}{W^*}$, where $W^*$ is the solution of following optimization problem:}{The first part follows in the exact same manner as in \cite[Lemma 1]{Marden}. Towards part (ii), let $\hatGG$ be the set of games that contains one  $\hat{G}$ for each $G \in \GG$ such that $\hat{G}$ matches $G$ in everything except the action sets of the agents.  The action set of any agent $i$ in $\hat{G}$ only contains the action at worst NE of $G$ and that at optimal action profile, that is, $\A_i = \{\ane_i,\aopt_i\}$. Then, repeating the exact arguments as in step 1 and 2 of the proof of \cite[Theorem 2]{Marden}, one can prove that the price of anarchy  $\poa= \frac{1}{W^*}$, where $W^*$ is the solution of following optimization problem:}
\begin{eqnarray}\label{eqn_gen_org_ot}
    W^* &=& \sup_{\hat{G}\in \hGG} W(\aopt)\nonumber\\
 \mbox{s.t.} && U_i(\ane) \ge U_i(\aopt_{i},\ane_{-i})\hspace{5mm} \forall \ \ i \in N,\nonumber\\
&& W(\ane)=1. 
\end{eqnarray}To construct a linear program solving optimization problem in \eqref{eqn_gen_org_ot} consider the following relaxation,
\begin{eqnarray}\label{eqn_gen_relax_ot}
    V^* &=& \sup_{\hat{G}\in \GG} W(\aopt)\nonumber\\
 \mbox{s.t.}&& \sum_{i \in \C_j} U_i(\ane) - U_i(\aopt_{i},\ane_{-i}) \ge 0,\hspace{5mm} \mbox{for } j \in [k],\nonumber\\
&& W(\ane)=1.
\end{eqnarray}Further, let
\begin{enumerate}[(a)]
    \item $x_{j,r} \in \{0,[\kappa_j]\}$ be the number of agents in $\C_j$ selecting resource $r$ in both $\ane$ and $\aopt$. 
    \item $a_{j,r} + x_{j,r} \in \{0,[\kappa_j]\}$ be the number of agents in $\C_j$ selecting resource $r$ in $\ane$.
    \item $b_{j,r}  + x_{j,r}  \in \{0,[\kappa_j]\}$ be the number of agents in $\C_j$ selecting resource $r$ in $\aopt$.
    \item  $A_{r}$ be total number of agents who select resource $r$ at $\ane$, thus $A_{r}:= \sum_{j \in [k]} a_{j,r} + x_{j,r}$.
    \item  $A_{r,j}$ be the number of agents from classes $\O_j$ who select resource $r$ at $\ane$, thus $A_{r,j} := \sum_{l \in \O_j} a_{l,r} + x_{l,r}$.
    \item  $B_{r}$ be total number of agents who select resource $r$ at $\aopt$ thus  $B_{r} := \sum_{j \in [k]} b_{j,r} + x_{j,r}$.
\end{enumerate}For each $t \in \I$ of \eqref{eqn_gen_I}, define $\R(t)$ as follows:
\begin{eqnarray}
    \R(t)&=& \{r \in \R: a_{j,r}+x_{j,r} = a_j+x_j,\ x_{j,r} = x_j\nonumber \\
    &&\hspace{10mm}b_{j,r}+x_{j,r} = b_j+x_j \mbox{ for } j\in[k]\}\label{eqn_Rt}
\end{eqnarray}to be the set of all the resources that are selected by exactly $a_j+x_j$ agents from $\C_j$ at $\ane$, $b_j+x_j$  at $\aopt$ and $x_j$ agents in both $\ane$ and $\aopt$. Further define $\theta(t)$ to be the sum of the values of the resources in $\R(t)$, i.e., $\theta(t) = \sum_{r \in \R(t)} v_r$. We can use $\{\theta\}$ to define the quantities in  \eqref{eqn_gen_relax_ot}. Let $A_t,B_t,A_{t,j}$ be as in \eqref{eqn_short}, then
\begin{eqnarray}\label{eqn_w_ane}
    W(\ane) &=&  \sum_{r\in \R} v_r \ w(A_r)\\
            &=& \sum_{t\in\I} w(A_t) \sum_{r \in \R(t)} v_r = \sum_{t\in\I} w(A_t) \theta(t).\nonumber
\end{eqnarray}Using similar logic, we get
\begin{eqnarray}\label{eqn_w_aopt}
    W(\aopt) &=& \sum_{t\in\I} w(B_t) \theta(t) \hspace{5mm} \mbox{ and } \\
    \sum_{i \in \C_j} U_i(\ane) 
                                &=& \sum_{t\in\I} (a_j+x_j)f_j(A_{t,j})\theta(t).\label{eqn_U_ane}
\end{eqnarray}Lastly, $\sum_{i \in \C_j} U_i(\aopt_{i},\ane_{-i})$ equals,
\begin{eqnarray}\label{eqn_U_ai_ane}
&&\hspace{-7mm}= \sum_{r \in \R} v_r[ x_{j,r}  f_j(A_{r,j})+ b_{j,r}  f_j(A_{r,j}+1)]\nonumber\\
&&\hspace{-7mm}= \sum_{t \in \I} [ x_j f_j(A_{t,j})+  b_j  f_j(A_{t,j}+1)] \theta(t).
\end{eqnarray}Thus, the optimization problem \eqref{eqn_gen_relax_ot} is equivalent to
\begin{eqnarray}\label{eqn_LP_with_sup}
   W^* &=& \sup_{\theta(t)}\sum_{t\in \I} w(B_t) \theta(t) \nonumber\\
\mbox{s.t.}\hspace{5mm}&& \hspace{-10mm} \sum_{t\in \I} [ a_j f_j(A_{t,j})-  b_j f_j(A_{t,j}+1)] \theta(t) \ge 0,\hspace{1mm} j \in [k],\nonumber\\
    && \hspace{-10mm}\sum_{t\in\I} w(A_t)\theta(t)=1,\nonumber\\
    && \hspace{-10mm} \theta(t) \ge 0 \ \hspace{5mm} \mbox{ for all }\ \ t\in \I, \\
  \mbox{and} && \hspace{-10mm}\ f_j(0)= f_j(|\N_j|+1) = w(0) =0, \ \forall \ j\in[k].\nonumber
\end{eqnarray}To see that the maximum in above is achieved, note that the objective function is continuous in $\{\theta\}$. All $\theta(t)\ge 0$ are bounded below. The constraint on $W(\ane)=1$ implies that $\theta(t)$ is bounded for all $t \in \I$ with $A_{t} \ge 1$. For $t \in \I$ such that $A_t= 0$, the equilibrium constraint for class $\C_j$ can be re-written as,
\begin{eqnarray*}
    \sum_{\substack{t\in\I\\ A_t=0}} b_j f_j(1) \theta(t) &\le& \sum_{\substack{t\in\I\\ A_t\ge 1}} [ a_j f_j(A_{t,j}-  b_j f_j(A_{t,j}+1)] \theta(t)
\end{eqnarray*}which provides the required boundedness as $f_j(1) >0$. Thus  $\{\theta(t)\}$ belong to a compact domain and optimization problem in \eqref{eqn_gen_relax_ot} is equivalent to the linear program in \eqref{eqn_poa_primal_lp_gen}.

It is easy to observe that $V^* \ge W^*$ as any feasible point for optimization problem \eqref{eqn_gen_org_ot} is also a feasible point for optimization problem \eqref{eqn_gen_relax_ot} and hence for \eqref{eqn_poa_primal_lp_gen}. Lemma \ref{lem_gen_relaxation_equiv} shows that $V^* \le W^*$, and the proof follows. \eop
\begin{lemma}\label{lem_G_func}
For any $j \in [k]$, let $G_j(\cdot)$ be defined as
\begin{eqnarray}\label{eqn_G_gen}
     G_j(\aa) &=& \sum_{r \in \cup_i a_i} v_r  \sum_{l=1}^{|\aa|_r^{\N_j}} f_j(l).
\end{eqnarray}Then, $  G_j(\aa) -G_j(b_i,a_{-i}) = U_i(\aa)-U_i(b_i,a_{-i})$  for all  $ b_i \in \A_i$, $i \in \C_j $ and all $j\in[k]$.
\end{lemma}
\noindent\textbf{Proof:} When an agent $i\in \C_j$ deviates to $b_i = \{a_i \backslash R_1\}\cup R_2$ where $R_1,R_2 \subset \R$, then 
\begin{eqnarray*}
   U_i(b_i,a_{-i}) = \sum_{r \in  a_i\backslash R_1} v_r f_j(|\aa|^{\N_j}_r) + \sum_{r \in  R_2} v_r f_j(|\aa|^{\N_j}_r+1)\\
                   = U_i(a_i,a_{-i}) -\sum_{r \in R_1} v_r f_j(|\aa|^{\N_j}_r) + \sum_{r \in  R_2} v_r f_j(|\aa|^{\N_j}_r+1).
\end{eqnarray*}Further, $G_j(b_i,a_{-i})$ equals,
\begin{eqnarray*}
    &=& G_j (\aa)-\sum_{r \in R_1} v_r f_j(|\aa|^{\N_j}_r) + \sum_{r \in  R_2} v_r f_j(|\aa|^{\N_j}_r+1).
\end{eqnarray*}Hence the proof.  \eop
\begin{lemma}\label{lem_gen_relaxation_equiv}
    Consider $W^*$ as in \eqref{eqn_gen_org_ot} and $V^*$ as in \eqref{eqn_poa_primal_lp_gen}. It holds that  $V^* \le W^*$. 
\end{lemma}
\noindent\textbf{Proof:} Let  $\{\theta(t)\}$ be any feasible solution of LP \eqref{eqn_poa_primal_lp_gen} with objective value $V$. We construct a game instance $G$ that satisfies the constraints of \eqref{eqn_gen_org_ot} with the same objective value. This will conclude $V^* \le W^*$. 

Let $\n := \prod_{j=1}^k \kappa_j$. For each $t\in \I$, construct  resources $r(t,q)$ for  $q\in[\n]$ with value $v_{r(t,q)}=\frac{\theta(t)}{\n}$. For each $j\in [k]$ and $t \in \I$ define the following sets,
\begin{eqnarray*}
    K_j(t,l) &=&  \cup_{q=(l-1) \frac{\n}{\kappa_j}+1}^{l \frac{\n}{\kappa_j}}\ \ r_{(t,q)}  \hspace{5mm}\mbox{ for } l=1,\dots, \kappa_j.
\end{eqnarray*}%
Add these sets to the action sets of agents in $\C_j$ in the exact same  manner as in proof of Lemma 3 in \cite{Marden}. Basically, order the set of agents $\C_j$ from $1$ to $\kappa_j$, and then for the agent at $i$-th position in $\C_j$, add the sets of resources to the equilibrium and optimal action as\footnote{With slight abuse of notation, we use $\ane_{ij}$ and $\aopt_{ij}$ to represent the action set of player at $i$-th position in $\C_j$, whereas this player may have been assigned a different label from $[n]$ in the original problem.}
\begin{eqnarray*}
   &&\hspace{-7mm} \ane_{ij}=
      \cup_{l=1}^{\kappa_j} \{ K_j(t,l)  \mbox{ s.t. } a_j+x_j \ge 1+ (l-i)\hspace{-3mm}\mod \kappa_j\},  \\ 
   &&\hspace{-7mm}   \aopt_{ij} \hspace{-2mm}=\hspace{-1mm}
      \cup_{l=1}^{\kappa_j}\hspace{-1mm} \{ K_j(t,l)  \mbox{ s.t. } b_j+x_j \ge 1+ (l-i+b_j)\hspace{-3mm}\mod \kappa_j\} .   
\end{eqnarray*}Informally, above construction implies that for fixed $t\in\I$ and $j \in \C_j$,  position $\kappa_j$ sets $K_j(t,l)$ on a circle indexed by $l \in [\kappa_j]$. Then, for the agent at $i$-th position in $\C_j$, equilibrium action  $\ane_{ij}$ contains $a_j+x_j$ sets starting from $ K_j(t,i)$ and moving clockwise. The optimal action $\aopt_{ij}$ contains $b_j+x_j$ sets starting from $ K_j(t,1+(i-1-b_j)\mod \kappa_j)$ and moving clockwise. The same process is repeated for all $t \in \I$ and all $\C_j$. This results in a game where for all $t \in \I$,
\begin{enumerate}[\textbf{B}.1]
\item  Any resource $r_{(t,q)}$ for $q \in[\n]$ has a value of $\frac{\theta(t)}{\n}$.
    \item  Any  $r_{(t,q)}$ for $q \in[\n]$ is selected by exactly $a_j+x_j$  agents from $\C_j$ at $\ane$, thus $A_t$ agents (see \eqref{eqn_short}). Further, $r_{(t,q)}$ is selected by $b_j+x_j$ agents from class $\C_j$ at $\aopt$ (thus $B_t$ agents).
    \item Each agent in $\C_j$ selects $(a_j+x_j)$ sets out of $\{K_j(t,l)\}_{l \in [\kappa_j]}$ at $\ane$, each of which contains $\frac{\n}{\kappa_j}$ resources. Thus every agent selects $\frac{\n}{\kappa_j}(a_j+x_j)$ resources at $\ane$. Every agent selects $b_j+x_j$ such sets  at $\aopt$, thus selects $\frac{\n}{\kappa_j}(b_j+x_j)$ resources. Further, $x_j$ sets, and hence $\frac{\n}{\kappa_j} x_j$ resources are common in $\ane$ and $\aopt$.

\end{enumerate}%
We now prove that $W(\ane)=1$ and $W(\aopt)=V$. Observe that 
\begin{eqnarray}\label{eqn_wane_1}
    W(\ane) = \hspace{-1mm}\sum_{t \in\I}  \sum_{q\in [\n]}\hspace{-1mm} v_{r(t,q)} =w(A_t)\sum_{t\in\I} \theta(t) w(A_t)=1\hspace{2mm}
\end{eqnarray}by the constraint in the LP in \eqref{eqn_poa_primal_lp_gen}. Similar arguments prove that $W(\aopt)=V$. 


For any $t \in \I$, the resource $r_{(t,q)}$ is selected by $A_{t,j}= \sum_{p \in \O_j} a_{p}+x_{p}$ agents from $\N_j$ for $j \in [k]$. The function \eqref{eqn_G_gen} of Lemma \ref{lem_G_func} at $\ane$ equals,
\begin{equation}\label{eqn_G_eq}
      G_j(\ane)=  \sum_{q \in [\n]} \sum_{t\in\I}  \frac{\theta(t)}{{\n}}\sum_{l=1}^{A_{t,j}}f_j(l) =\frac{1}{{\n}}\sum_{t\in\I} {\n}  \theta(t)\sum_{l=1}^{A_{t,j}}f_j(l).
\end{equation}When an agent $i \in \C_j$ unilaterally deviates from $\ane$ to $(\aopt_i,\ane_{-i})$, then each $K_j(t,l) \in \ane_i$ which is not in $\aopt_i$ would be selected one less time, and each $K_j(t,l) \in \aopt_i$ which is not in $\ane_i$ would be selected one more time.  Thus,  $a_j$ sets of the type $K_j(t,l)$ get selected by one less agent and $b_j$ sets of the type $K_j(t,l)$ get selected by one extra agent. Thus for each $t \in\I$, $\frac{\n}{\kappa_j}a_j$ resources get selected by $A_{t,j} -1$ agents from $\N_j$ and $\frac{\n}{\kappa_j}b_j$ resources get selected by $A_{t,j}+1$ agents from $\N_j$, rest are selected by $A_{t,j}$ agents from $\N_j$. Then $G_j(\aopt_i,\ane_{-i})$ equals,
\begin{eqnarray}\label{eqn_G_devi}
    &&\hspace{-0.75cm}=\ \frac{1}{{\n}}\sum_{t\in\I}  \theta(t)\Big[\frac{\n b_j}{\kappa_j}\sum_{l=1}^{A_{t,j}+1}f_j(l)+\frac{\n a_j}{\kappa_j} \sum_{l=1}^{A_{t,j}-1}f_j(l)\nonumber\\
    &&\hspace{2.5cm}+\Big({\n}-\frac{\n a_j}{\kappa_j} -\frac{\n b_j}{\kappa_j}\Big)\sum_{l=1}^{A_{t,j}}f_j(l)\Big]\\
   &&\hspace{-0.75cm}=\ \frac{1}{{\n}}\sum_{t\in\I}  \theta(t)\Big[\frac{\n b_jf_j(A_{t,j}+1)}{\kappa_j} - \frac{\n a_j f_j(A_{t,j})}{\kappa_j}+{\n}\sum_{l=1}^{A_{t,j}}f_j(l)\Big].\nonumber
\end{eqnarray}From \eqref{eqn_G_eq} and \eqref{eqn_G_devi}, $ G_j(\ane)-G_j(\aopt_i,\ane_{-i}) $ equals
\begin{eqnarray}\label{eqn_proving_eq}
    &&\hspace{-0.75cm} =\frac{1}{{\kappa_j}}\sum_{t\in\I}  \theta(t) [a_j f_j(A_{t,j}) -b_jf_j(A_{t,j}+1)] \ \ge\ 0
\end{eqnarray}by the constraint in the LP \eqref{eqn_poa_primal_lp_gen}.  Similar arguments follow for all $j\in[k]$. By Lemma \ref{lem_G_func}, $\ane$ is a Nash equilibrium, hence satisfies the constraints in \eqref{eqn_gen_org_ot}, hence the proof.  \eop

\section*{Proof of Theorem \ref{thm_dual_lp_gen}}

The dual of the LP in \eqref{eqn_poa_primal_lp_gen} equals (see  e.g.,\cite{Bertsimas})
\begin{eqnarray}\label{eqn_poa_dual_lp_gen_temp}
V^* & =&  \min_{\lambda_1,\dots,\lambda_k,\mu} \ \   \mu \hspace{5mm} \mbox{ subject to }\\
    && \hspace{-15mm} w(B_t) + \sum_{j \in [k]} \lambda_j  [ a_j f_j(A_{t,j})-  b_j f_j(A_{t,j}+1)]  \le  \mu  w(A_t)\nonumber \\
    && \hspace{5cm}\ t  \in \I_\R,\nonumber \\
    && \hspace{-15mm}\mu \in \mathbb{R}, \lambda_j \ge 0, f_j(0)= f_j(|\N_j|+1) = w(0) =0,  \forall \ j\in[k]. \nonumber
\end{eqnarray} By strong duality  (\cite{Bertsimas}), $V^*$ of above equals $W^*$ of \eqref{eqn_poa_primal_lp_gen}. It remains to show that considering all $t \in \I$ of \eqref{eqn_gen_I} is equivalent to considering all $t \in \I_\R$ of \eqref{eqn_gen_IR}. 

Observe that $t \in \I$ such that $A_t=0$ or $B_t=0$ are all contained in $\I_\R$. Next, we show that the constraints corresponding to $t\in \I$  such that $A_t \ne 0$ and $B_t \ne 0$ are equivalent to the constraints corresponding to  $t\in \I_\R$. 

Consider a change of coordinates from $t$ to $\Tilde{t}=(\tltuple)$ where $\Tilde{t}_j$ equals $(l_j,x_j,m_j)$ with $l_j := a_j+x_j$ and $m_j := b_j+x_j$. Then the constraint in \eqref{eqn_poa_dual_lp_gen_temp} can be re-written as,
\begin{eqnarray}\label{eqn_gen_cons}
   \mu  w(A_t)
                             &\ge&w(B_t)  + \sum_{j\in[k]} \lambda_j  [l_j f_j(A_{t,j})- m_j f_j(A_{t,j}+1)\nonumber\\
                             &&\hspace{1.5cm}+ x_j[f_j(A_{t,j}+1) -f_j(A_{t,j})]]. \hspace{2mm}
\end{eqnarray}for all $\Tilde{t} \in \hat{\I}$ where $  \hat{\I}$ equals,
\begin{eqnarray*}
   &&\hspace{-0.75cm}\Big\{(\tltuple) : (l_j,x_j,m_j) \in \mathbb{N}^3_{\ge 0},\ 0\le l_j-x_j+m_j\le\kappa_j, \\
   &&l_j \ge x_j,\  m_j \ge x_j, \  \sum_{j \in[k]}l_j \ne 0,\  \sum_{j \in[k]}m_j \ne 0\nonumber \\
    &&\hspace{2mm}\mbox{ and }\ 1 \ \le\ \sum_{j \in[k]} (l_j-x_j+m_j)  \le \ n\Big\}.
\end{eqnarray*}In the remaining of this proof, fix $l_j$ for all $j$, and let $x_j$ and $m_j$ move freely in $\hat{\I}$. When $l_j = \kappa_j$ for all $j\in[k]$, then $m_j = x_j$ since $m_j-x_j \ge 0$ and $-x_j+m_j \le 0$.  This implies $b_j = 0$ for all $j$, and note that this point is also contained in $\I_\R$. Now consider the $\Tilde{t} \in\hat{\I}$ such that $l_j \ne \kappa_j$ for at least one $j\in[k]$, and define two sets $\R_1$ and $\R_2$ as,
\begin{eqnarray*}
    \R_1 &=& \{j \in [k]:\ f_j(A_{t,j}+1) -f_j(A_{t,j}) \le 0\},\\
    \R_2 &=& \{j \in [k]:\ f_j(A_{t,j}+1) -f_j(A_{t,j}) > 0\}.
\end{eqnarray*}For any fixed $l_j,m_j$ for $j\in [k]$, the most binding constraint in \eqref{eqn_gen_cons} arises when $x_j$ is picked as small as possible for $j \in \R_1$ (since $x_j \ge 0$ and is multiplied with a negative coefficient), and $x_j$ is picked as large as possible for $j \in \R_2$ (since $x_j \ge 0$ and is multiplied with a positive coefficient). 

For $j \in \R_1$, since $x_j \ge l_j + m_j - \kappa_j$, for fixed $m_j$ and $l_j$, the smallest value $x_j$ can take is $x_j =\max\{0,l_j+m_j-\kappa_j\}$. There are two possibilities,
\begin{enumerate}[\textbf{B}.1]
    \item when $l_j+m_j \le \kappa_j$, that is when $a_j+b_j+2x_j \le n$, then $x_j =0$ implying $a_j.x_j.b_j =0$,
    \item when $l_j+m_j > \kappa_j$, that is when $a_j+b_j+2x_j >\kappa_j$ then $x_j = l_j+m_j - \kappa_j$ implying $a_j+x_j+b_j = \kappa_j$.
\end{enumerate}Similarly, for $j \in \R_2$, since $x_j \le l_j $ and $x_j \le m_j$, for fixed $m_j$ and $l_j$ largest value of $x_j$ is $x =\min\{l_j,m_j\}$. Again, there are two possibilities,
\begin{enumerate}[\textbf{B}.1]
\setcounter{enumi}{2}
    \item when $l_j \le m_j$, that is when $a_j\le b_j$ then $x_j =l_j$ and thus $a_j =0$ implying $a_j.x_j.b_j =0$,
    \item when $l_j > m_j$, that is when $a_j > b_j$ then $x_j =m_j$ and thus $b_j =0$ implying $a_j.x_j.b_j =0$.
\end{enumerate}In all, for any $j \in [k]$, one of the \textbf{B}.1-\textbf{B}.4 holds at the most binding constraint. Since all such quantities are included in $\I_\R$, it is sufficient to consider $\I_\R$ and LP in \eqref{eqn_poa_dual_lp_gen_temp} is equivalent to the LP \eqref{eqn_poa_dual_lp_gen}. \eop

\TR{\section{Proof of Theorem \ref{thm_opt_lp_gen}}
From Theorem \ref{thm_primal_lp_gen}, if $f_j(1)\le 0$ for any $j \in [k]$, then the resulting price of anarchy is zero. While if $f_j(1)>0$ for all $j$, by Theorem \ref{thm_primal_lp_gen}, $PoA(f,w,n)>0$, hence $f$ with $f_j(1)\le 0$ for any $j \in [k]$ can not be optimal.   From the constraint in (primal) LP \eqref{eqn_poa_primal_lp_gen}, it is clear that scaling $f_j$ with a positive constant does not change the PoA, thus, without loss of generality, we only  consider $f_j \in F_j$ where
$$F_j = \{f_j:[|\N_j|] \to \mathbb{R} \mbox{ s.t. } f_j(1)\ge w(1)\}.$$
When $f_j \in F_j$ for all $j$, the price of anarchy can be computed using \eqref{eqn_poa_dual_lp_gen}. Thus, the problem of devising utility generating mechanisms that maximize the price of anarchy is same as devising the ones that minimize $W^*$ of \eqref{eqn_poa_dual_lp_gen}. That is,
\begin{eqnarray}\label{eqn_poa_opt_temp_lp_gen}
\arg\min_{\{f_j \in F_j\}}&& \min_{\lambda_1,\dots,\lambda_k,\mu} \ \   \mu\\
     &&\hspace{-2cm}\mbox{s.t.} \hspace{5mm}w\Big(\sum_{j \in [k]} b_j + x_j\Big)  - \mu  w\Big(\sum_{j \in [k]} a_j + x_j\Big) + \sum_{j \in [k]} \lambda_j  \Big[ a_j f_j\Big(\sum_{p \in \O_j}a_p+x_p \Big)-  b_j f_j\Big(1+\sum_{p \in \O_j}a_p+x_p \Big)\Big]  \ \le \ 0 \nonumber \\
    && \hspace{10cm} \forall (\ttuple) \  \in\  \I_\R \nonumber\\
    && \lambda_1, \dots \lambda_k\ \ge \ 0,\ \mu \mbox{ is unrestricted and } f_j(0)= f_j(|\N_j|+1) = w(0) =0 \mbox{ for all }j\in[k] \nonumber
\end{eqnarray}Lemma \ref{lem_bdd_opt_f} shows that the above program is well posed, that is, the minimum is achieved by $f_j$ with bounded components for all $j \in [k]$. Although this is a non-linear problem, $\lambda_j$ and $f_j(\cdot)$ are always multiplied together. Hence, define a new variable $\f_j(i):= \lambda_j f_j(i)$ for all $i \in \{0,[|\N_j|+1]\}$ and all $j \in [k]$. %
For any $j \in [k]$, and $(\ttuple)\in \I_\R$ such that $a_p,x_p = 0$ for all $p \in [k]$, $b_j=1$ and $b_p=0$ for all $p \in [k]\backslash j$, the constraint in \eqref{eqn_poa_opt_temp_lp_gen} gives $\f_j(1) = \lambda_j f_j(1) \ge w(1)$ thus $\f_j \in F_j$. This also implies $\lambda_j \ge w(1)/f_j(1)>0$ for all $j$ since $w(1)>0$ and $f_j(1)>0$. Folding the $\min$ operator gives
\begin{eqnarray}\label{eqn_poa_opt_tempp_lp_gen}
(\f_1^*,\dots,\f_k^*,\mu^*)  & \in & \min_{\lambda_1,\dots,\lambda_k,\mu} \ \   \mu\\
     &&\hspace{-2cm}\mbox{s.t.} \hspace{5mm}w\Big(\sum_{j \in [k]} b_j + x_j\Big)  - \mu  w\Big(\sum_{j \in [k]} a_j + x_j\Big) + \sum_{j \in [k]} \Big[ a_j \f_j\Big(\sum_{p \in \O_j}a_p+x_p \Big)-  b_j \f_j\Big(1+\sum_{p \in \O_j}a_p+x_p \Big) \Big] \ \le \ 0 \nonumber \\
    && \hspace{10cm} \forall (\ttuple) \  \in\  \I_\R \nonumber\\
    &&\mu \mbox{ is unrestricted and } \f_j(0)= \f_j(|\N_j|+1) = w(0) =0 \mbox{ for all }j\in[k]. \nonumber
\end{eqnarray}Finally observe that $\f_j^*$ is feasible for original program since $f_j^*(1) \ge w(1)$ for all $j$. Further, the PoA given by $\f_1^*,\dots,\f_k^*$ is same as the one given by $\f_1^*,\dots,\f_k^*$ since $\f_j^* = \lambda^*_j f_j^*$ and scaling by a positive constant does not change the PoA. Thus  $\f_1^*,\dots,\f_k^*$ solving \eqref{eqn_poa_opt_tempp_lp_gen} must be optimal and optimal PoA equals $1/\mu^*$. \eop

\begin{lemma}\label{lem_bdd_opt_f}
    The minimum appearing in \eqref{eqn_poa_opt_temp_lp_gen} is attained by $\{f_j\}$ with bounded components.
\end{lemma}
\noindent\textbf{Proof:} In the following, we show that the infimum 
\begin{eqnarray}\label{eqn_poa_opt_temp_lp_gen_inf}
\inf_{\{f_j \in F_j\}}&& \min_{\lambda_1,\dots,\lambda_k,\mu} \ \   \mu\\
     &&\hspace{-2cm}\mbox{s.t.} \hspace{5mm}w\Big(\sum_{j \in [k]} b_j + x_j\Big)  - \mu  w\Big(\sum_{j \in [k]} a_j + x_j\Big) + \sum_{j \in [k]} \Big[ a_j f_j\Big(\sum_{p \in \O_j}a_p+x_p \Big)-  b_j f_j\Big(1+\sum_{p \in \O_j}a_p+x_p \Big)\Big]  \ \le \ 0 \nonumber \\
    && \hspace{10cm} \forall (\ttuple) \  \in\  \I_\R \nonumber\\
    && \lambda_1, \dots \lambda_k\ \ge \ 0,\ \mu \mbox{ is unrestricted and } f_j(0)= f_j(|\N_j|+1) = w(0) =0 \mbox{ for all }j\in[k] \nonumber
\end{eqnarray}is achieved by $\{f_j\}$ with bounded components. Since the price of anarchy of mechanism $\{f_j\}$ and $\{\alpha_j f_j\}$ is same for $\alpha_j >0$, the above problem is equivalent to
\begin{eqnarray}\label{eqn_poa_opt_tempp_lp_gen_inf}
\inf_{\substack{\{f_j \in F_j\}\\ f_j(1) = w(1)}}&& \min_{\lambda_1,\dots,\lambda_k,\mu} \ \   \mu\\
     &&\hspace{-2cm}\mbox{s.t.} \hspace{5mm}w\Big(\sum_{j \in [k]} b_j + x_j\Big)  - \mu  w\Big(\sum_{j \in [k]} a_j + x_j\Big) + \sum_{j \in [k]} \lambda_j  \Big[ a_j f_j\Big(\sum_{p \in \O_j}a_p+x_p \Big)-  b_j f_j\Big(1+\sum_{p \in \O_j}a_p+x_p \Big)\Big]  \ \le \ 0 \nonumber \\
    && \hspace{10cm} \forall (\ttuple) \  \in\  \I_\R \nonumber\\
    && \lambda_1, \dots \lambda_k\ \ge \ 0,\ \mu \mbox{ is unrestricted and } f_j(0)= f_j(|\N_j|+1) = w(0) =0 \mbox{ for all }j\in[k] \nonumber
\end{eqnarray}Choose $f_p \in F_p$ for all $p \in[k]$. Fix $j \in [k]$ and construct a mechanism $f_{j,M}$ from $f_j$ as follows: let $f_{j,M}(l)=M$ for $M \in \mathbb{R}$ for a fixed $l \in [2,|\N_j|]$, while $f_{j,M}$ matches $f_j$ for the remaining components\footnote{For $j$ with $|\N_j|=1$, the utility generating mechanism is given by only $f_j(1)$. Thus, scaling this by any positive constant does not change price of anarchy. By the same theorem, if $f_j(1)$ is scaled by a negative constant, the price of anarchy becomes zero and can not be optimal. Thus, the optimal mechanism contains $f_j(1)=w(1)$ for $j$ with $|N_j|=1$.}. To avoid any confusion, we represent $\{f_1,\dots,f_k\}=(f_{j},f_{-j})$ explicitly in the PoA expression.  We show that there exists $M^+$ and $M^-$ such that $PoA((f_{j,M},f_{-j}),w,n) \le PoA ((f_{j},f_{-j}),w,n)$ for all $M\le M^-$ and all $M \ge M^+$. This concludes that the optimal can not be achieved by $f_j$ with unbounded components in a single direction for any $j \in [k]$. Identical arguments can show that optimal can not be achieved by $f_j$ with unbounded components in two or more directions.

From Theorem \ref{thm_primal_lp_gen}, the price of anarchy $PoA((f_{j},f_{-j}),w,n) = 1/W^*$ where $W^*$ is the solution of primal LP \eqref{eqn_poa_primal_lp_gen} and is finite. Thus  $PoA((f_{j},f_{-j}),w,n)>\epsilon$ for some $\epsilon>0$. Now, consider $(\ttuple) \in \I_\R$ such that $a_p =x_p =0$ for all $p \in [k]$, $b_j =1$ and $b_p=0$ for all $p \ne j$. Then, from  the constraint in \eqref{eqn_poa_opt_tempp_lp_gen_inf} any feasible $\lambda_j$ satisfies,
\begin{eqnarray*}
  \lambda_j f_{j,M}(1)   &\ge& w(1) \\
  \lambda_j   &\ge& \frac{w(1)}{ f_{j,M}(1)} \ =\ 1 \\
\end{eqnarray*}Further for $(\ttuple) \in \I_\R$ such that $b_p =x_p =0$ for all $p \in [k]$, $a_j = l \ge 2$ and $a_p=0$ for all $p \ne j$, the constraint becomes,
\begin{eqnarray*}
    \mu\ w(l) &\ge& \lambda_j \ l \ f_{j,M}(l) \ \ge \ l f_{j,M} \ = \ l\ M
\end{eqnarray*}thus, 
\begin{eqnarray*}
    PoA((f_{j,M},f_{-j}),w,n) & =& \frac{1}{\mu^*} \ \le \ \frac{1}{l\ M}
\end{eqnarray*}Thus, there exists some $M^+$ such that  $PoA((f_{j,M},f_{-j}),w,n)<\epsilon <PoA((f_{j},f_{-j}),w,n)$ for all  $M > M^+$.

For $M<0$, consider  $(\ttuple) \in \I_\R$ such that $a_p =0$ for all $p \in [k]$, $b_j =1$, $x_j = l-1 \ge 1$ and $b_p,\ x_p=0$ for all $p \ne j$, the constraint becomes,
\begin{eqnarray*}
  \mu & \ge &  \frac{w(l) - \lambda_j f_j( l)}{ w(l-1) } \ = \ \frac{w(l) }{ w(l-1) } -\frac{ \lambda_j M}{ w(l-1) } \ \ge \ \frac{w(l) }{ w(l-1) } -\frac{  M}{ w(l-1) }   \\
\end{eqnarray*}where the last inequality holds since $M<0$ and $\lambda_j \ge 1$. Thus, using similar arguments as above, there exists some $M^-$ such that  $PoA((f_{j,M},f_{-j}),w,n)<\epsilon <PoA((f_{j},f_{-j}),w,n)$ for all  $M < M^-$.}{}
{\section*{Proof of Theorem \ref{thm_opt_lp_blind}}}
The proof of this theorem follows in similar lines as the proof of Theorem \ref{thm_primal_lp_gen} - Theorem \ref{thm_opt_lp_gen}. Thus, we briefly provide the steps that differ from the original. Part (i) holds exactly as in proof of Theorem \ref{thm_primal_lp_gen}.  Towards part (ii), as discussed in the proof of Theorem \ref{thm_primal_lp_gen}, \eqref{eqn_gen_org_ot} holds even for this case. Then, one can rewrite \eqref{eqn_gen_relax_ot} for classes $\C_1$ and $\C_2$. Then one can define all the quantities in (a)-(f)  in the same manner except for $A_{r,1}$ corresponding to class of blind agents, which will also not be required in the proof. Then, one can define $\I$ and $\R(t)$ of \eqref{eqn_Rt} in the exact same manner. It is easy to see that \eqref{eqn_w_ane}-\eqref{eqn_w_aopt} follow. Further, \eqref{eqn_U_ane}-\eqref{eqn_U_ai_ane} hold for class $\C_2$, and for class $\C_1$, $ \sum_{i \in \C_1} U_i(\ane) = \sum_{t\in\I} (a_1+x_1)f_{bl}(1)\theta(t)$, and
\begin{eqnarray}
    \sum_{i \in \C_1} U_i(\aopt_i,\ane_{-i}) &=& \sum_{t\in\I} (x_1+b_1)\fbl(1)\theta(t).\label{eqn_u_aopt_ane_bl}
\end{eqnarray}Thus, the problem in \eqref{eqn_gen_relax_ot} is equivalent to problem in \eqref{eqn_LP_with_sup} for blind agents case with the modified equilibrium constraint for $\C_1$ modified as $\sum_{t\in\I} (a_1-b_1) \fbl(1) \theta(t) \ge 0$ which equals
\begin{eqnarray}\label{eqn_bl_constraint}
    \sum_{t\in\I} (a_1-b_1) \theta(t) \ \ge 0.
\end{eqnarray}The arguments for maximum being achieved follow in the exact same manner. Thus, \eqref{eqn_gen_relax_ot} is equivalent to the LP in  \eqref{eqn_poa_primal_lp_gen} with constraint for $\C_1$ changed to \eqref{eqn_bl_constraint}. To show the equivalence of the problems in \eqref{eqn_gen_relax_ot} and \eqref{eqn_gen_org_ot} for this case, construct a game exactly as in proof of Lemma \ref{lem_gen_relaxation_equiv} and note that \eqref{eqn_wane_1} follows. Further \eqref{eqn_G_eq} follows for class $\C_2$ of normal agents. Let $|\aa|^b_r$ be the number of blind agents selecting resource $r$ in action profile $\aa$, and redefine $G_1(\aa)$ for class of blind agents as $G_1(\aa)= \sum_{r \in \R} v_r |\aa|^b_r$, then $ G_1(\ane) =  \sum_{t\in\I} (a_1+x_1) \theta(t)$. 
It is easy to verify that for any blind agent $i \in \C_1$, Lemma \ref{lem_G_func} holds. When $i \in \C_1$ deviates to $\aopt_i$, then $ G_1(\aopt_i,\ane_{-i})$ equals $\frac{1}{\kappa} \sum_{t \in \I} (b_1-a_1)\theta(t) \ge 0$ by the constraint in LP and rest of the arguments follow. Thus \eqref{eqn_gen_relax_ot} and \eqref{eqn_gen_org_ot} are equivalent for blind agent case and two classes. This proves Theorem \ref{thm_primal_lp_gen} for this case with the equilibrium constraint for class  of blind agents $\C_1$ modified to \eqref{eqn_bl_constraint}. Then, proofs of Theorem \ref{thm_dual_lp_gen} and Theorem \ref{thm_opt_lp_gen} follow in a similar manner and lead to Theorem \ref{thm_opt_lp_blind}. \eop

\addtolength{\textheight}{-12cm}   

\end{document}